\documentclass[aps,prd]{revtex4}
\usepackage{docs}
\usepackage{graphicx}
\usepackage{dcolumn}
\usepackage{bm}

\begin{document}
\title{Fundamental limitations on ``warp drive'' spacetimes}%

\author{Francisco S. N. Lobo}%
\email{flobo@cosmo.fis.fc.ul.pt}
\affiliation{Centro de Astronomia
e Astrof\'{\i}sica da Universidade de Lisboa,\\
Campo Grande, Ed. C8 1749-016 Lisboa, Portugal}

\author{Matt Visser}%
\email{matt.visser@mcs.vuw.ac.nz}
\affiliation{School of Mathematical and Computing Sciences, \\
Victoria University of Wellington, \\
P.O.Box 600, Wellington, New Zealand}

\begin{abstract}

  ``Warp drive'' spacetimes are useful as ``gedanken-experiments'' that
  force us to confront the foundations of general relativity, and
  among other things, to precisely formulate the notion of
  ``superluminal'' communication.  After carefully formulating the
  Alcubierre and Nat\'{a}rio warp drive spacetimes, and verifying their
  non-perturbative violation of the classical energy conditions, we
  consider a more modest question and apply linearized gravity to the
  weak-field warp drive, testing the energy conditions to first and
  second order of the warp-bubble velocity, $v$.  Since we take the
  warp-bubble velocity to be non-relativistic, $v \ll c$, we are not
  primarily interested in the ``superluminal'' features of the warp
  drive. Instead we focus on a secondary feature of the warp drive
  that has not previously been remarked upon --- the warp drive (if it
  could be built) would be an example of a ``reaction-less drive''.
  For both the Alcubierre and Nat\'{a}rio warp drives we find that the
  occurrence of  significant energy condition violations is not just
  a high-speed effect, but that the violations persist even at
  arbitrarily low speeds.

  A particularly interesting feature of this construction is that it is now
  meaningful to think of placing a finite mass spaceship at the centre
  of the warp bubble, and then see how the energy in the warp field
  compares with the mass-energy of the spaceship. There is no hope of
  doing this in Alcubierre's original version of the warp-field, since
  by definition the point in the centre of the warp bubble moves on a
  geodesic and is ``massless''.  That is, in Alcubierre's original
  formalism and in the Nat\'{a}rio formalism the spaceship is always
  treated as a test particle, while in the linearized theory we can
  treat the spaceship as a finite mass object. For both the Alcubierre
  and Nat\'{a}rio warp drives we find that even at low speeds the net
  (negative) energy stored in the warp fields must be a significant
  fraction of the mass of the spaceship.

\end{abstract}
\maketitle

\def\d{{\mathrm{d}}}

\def\ii{{\hat\imath}}
\def\jj{{\hat\jmath}}
\def\kk{{\hat k}}
\def\lll{{\hat l}}
\def\tt{{\hat t}}
\def\xx{{\hat x}}
\def\yy{{\hat y}}
\def\zz{{\hat z}}

\section{Introduction}

Alcubierre demonstrated that it is theoretically possible, within
the framework of general relativity, to attain arbitrarily large
velocities~\cite{Alcubierre}. In fact, numerous solutions to the
Einstein field equations are now known that allow ``effective''
superluminal travel
\cite{Natario,Morris,examples,surgery,Visser,Krasnikov,Everett,LLQ,Lemos,LoboSLT}.
Despite the use of the term superluminal, it is not ``really''
possible to travel faster than light, in any \emph{local} sense.
Providing a general \emph{global} definition of superluminal
travel is no trivial matter~\cite{VB,VBL}, but it is clear that
the spacetimes that allow ``effective'' superluminal travel
generically suffer from the severe drawback that they also involve
significant negative energy densities. More precisely,
superluminal effects are associated with the presence of
\emph{exotic} matter, that is, matter that violates the null
energy condition [NEC]. In fact, superluminal spacetimes violate
all the known energy conditions, and Ken Olum demonstrated that
negative energy densities and superluminal travel are intimately
related~\cite{Olum}. Although most classical forms of matter are
thought to obey the energy conditions, they are certainly violated
by certain quantum fields~\cite{twilight}. Additionally, certain
classical systems (such as non-minimally coupled scalar fields)
have been found that violate the null and the weak energy
conditions~\cite{B&V,BV1}. Finally, we mention that recent
observations in cosmology strongly suggest that the cosmological
fluid violates the strong energy condition [SEC], and provides
tantalizing hints that the NEC \emph{might} possibly be violated
in a classical regime~\cite{Riess,jerk,rip}.

For warp drive spacetimes, by using the ``quantum inequality''
deduced by Ford and Roman~\cite{F&R1}, it was soon verified that
enormous amounts of energy are needed to sustain superluminal warp
drive spacetimes~\cite{Roman,PfenningF}. To reduce the enormous
amounts of exotic matter needed in the superluminal warp drive,
van den Broeck proposed a slight modification of the Alcubierre
metric which considerably ameliorates the conditions of the
solution~\cite{Broeck1}. It is also interesting to note that, by using the
``quantum  inequality'', enormous quantities of negative energy densities
are needed to support the superluminal Krasnikov tube~\cite{Everett}.
However, Gravel and Plante~\cite{Gravel1,Gravel2} in a way similar
in spirit to the van den Broeck analysis, showed that it is theoretically
possible to lower significantly the mass of the Krasnikov tube.

In counterpoint, in this article we shall be interested in
applying linearized gravity to warp drive spacetimes, testing the
energy conditions at first and second order of the warp-bubble
velocity. We will take the bubble velocity to be non-relativistic,
$v\ll 1$. Thus we are not focussing attention on the
``superluminal'' aspects of the warp bubble, such as the
appearance of horizons~\cite{Hiscock,Clark,Gonz}, and of closed
timelike curves~\cite{EverettCTC}, but rather on a secondary
unremarked effect: The warp drive (\emph{if it can be realised in
nature}) appears to be an example of a ``reaction-less drive''
wherein the warp bubble moves by interacting with the geometry of
spacetime instead of expending reaction mass.

A particularly interesting aspect of this construction is that one
may place a finite mass spaceship at the origin and consequently
analyze how the warp field compares with the mass-energy of the
spaceship. This is not possible in the usual finite-strength warp
field, since by definition the point in the center of the warp
bubble moves along a geodesic and is ``massless''. That is, in the
usual formalism the spaceship is always treated as a test
particle, while in the linearized theory we can treat the
spaceship as a finite mass object.

Because we do not make any \emph{a priori} assumptions as to the
ultimate source of the energy condition violations, we will not
use (or need) the quantum inequalities. This means that the
restrictions we derive on warp drive spacetimes are more generic
than those derived using the quantum inequalities ---
the restrictions derived in this article will hold regardless
of whether the warp drive is assumed to be classical or quantum
in its operation.

We do not mean to suggest that such a ``reaction-less drive'' is
achievable with current technology --- indeed the analysis below
will, even in the weak-field limit, place very stringent
conditions on the warp bubble. These conditions are so stringent
that it appears unlikely that the ``warp drive'' will ever prove
technologically useful. The Alcubierre ``warp drive'', and the
related Nat\'{a}rio ``warp drive'', are likely to retain their
status as useful ``gedanken-experiments'' --- they are useful
primarily as a theoretician's probe of the foundations of general
relativity, and we wish to sound a strong cautionary note against
over-enthusiastic mis-interpretation of the technological
situation.

\section{Warp drive basics}

Within the framework of general relativity, Alcubierre demonstrated
that it is in principle possible to warp spacetime in a small {\it
  bubble-like} region, in such a way that the bubble may attain
arbitrarily large velocities. Inspired by the inflationary phase
of the early universe, the enormous speed of separation arises
from the expansion of spacetime itself. The simplest model for
hyper-fast travel is to create a local distortion of spacetime,
producing an expansion behind the bubble, and an opposite
contraction ahead of it. Nat\'{a}rio's version of the warp drive
dispensed with the need for expansion at the cost of introducing a
slightly more complicated metric.

The warp drive spacetime metric, in cartesian coordinates, is given by
(with $G=c=1$)
\begin{equation}
\d s^2=-\d t^2+ [\d\vec x - \vec \beta(x,y,z-z_0(t)) \; \d t]\cdot
[\d\vec x - \vec \beta(x,y,z-z_0(t)) \; \d t]\,.
\label{Cartesianwarpmetric-general}
\end{equation}
In terms of the well-known ADM formalism this corresponds to a
spacetime wherein \emph{space} is flat, while the ``lapse
function'' is identically unity, and the only non-trivial
structure lies in the ``shift vector'' $\beta(t,\vec x)$. Thus
warp drive spacetimes can also be viewed as specific examples of
``shift-only'' spacetimes. The Alcubierre warp drive corresponds
to taking the shift vector to always lie in the direction of
motion
\begin{equation}
\vec \beta(x,y,z-z_0(t)) =  v(t) \; \hat z \; f(x,y,z-z_0(t)),
\end{equation}
in which $v(t)=dz_0(t)/dt$ is the velocity of the warp bubble,
moving along the positive $z$-axis, whereas in the Nat\'{a}rio
warp drive the shift vector is constrained by being
divergence-free
\begin{equation}
\nabla \cdot \vec \beta(x,y,z) =  0.
\end{equation}

\subsection{Alcubierre warp drive}

In the Alcubierre warp drive the spacetime metric is
\begin{equation}
\d s^2=-\d t^2+\d x^2+\d y^2+\left[\d z-v(t)\;f(x,y,z-z_0(t))\; \d t\right]^2
\label{Cartesianwarpmetric}\,.
\end{equation}
The form function $f(x,y,z)$ possesses the general features of having
the value $f=0$ in the exterior and $f=1$ in the interior of the
bubble.  The general class of form functions, $f(x,y,z)$, chosen by
Alcubierre was spherically symmetric: $f(r)$ with
$r=\sqrt{x^2+y^2+z^2}$. Then
\begin{equation}
f(x,y,z-z_0(t)) = f(r(t))
\label{Alcubierreformfunction}
\qquad \hbox{with} \qquad
r(t)=\left\{[(z-z_{0}(t)]^2+x^2+y^2\right\}^{1/2}.
\end{equation}

Whenever a more specific example is required we adopt
\begin{equation}
f(r)=\frac{\tanh\left[\sigma(r+R)\right]
-\tanh\left[\sigma(r-R)\right]}{2\tanh(\sigma R)}\,,
\label{E:form}
\end{equation}
in which $R>0$ and $\sigma>0$ are two arbitrary parameters. $R$ is the
``radius'' of the warp-bubble, and $\sigma$ can be interpreted as
being inversely proportional to the bubble wall thickness. If $\sigma$
is large, the form function rapidly approaches a {\it top hat}
function, i.e.,
\begin{equation}
\lim_{\sigma \rightarrow \infty} f(r)=\left\{ \begin{array}{ll}
1, & {\rm if}\; r\in[0,R],\\
0, & {\rm if}\; r\in(R,\infty).
\end{array}
\right.
\end{equation}

It can be shown that observers with the four velocity
\begin{equation}
U^{\mu}=\left(1,0,0,vf\right),
\qquad\qquad
U_{\mu}=\left(-1,0,0,0\right).
\end{equation}
move along geodesics, as their $4$-acceleration is zero,
\emph{i.e.}, $a^{\mu} = U^{\nu}\; U^{\mu}{}_{;\nu}=0$. These
observers are called Eulerian observers. Their four-velocity is
normal to the spatial hypersurface with $t={\rm const}$, and they
are close analogues of the usual Eulerian observers of fluid
mechanics, which move with the flow of the fluid. The spaceship,
which in the original formulation is treated as a test particle
which moves along the curve $z=z_0(t)$, can easily be seen to
always move along a timelike curve, regardless of the value of
$v(t)$. One can also verify that the proper time along this curve
equals the coordinate time, by simply substituting $z=z_0(t)$ in
equation (\ref{Cartesianwarpmetric}). This reduces to $\d\tau=\d
t$, taking into account $\d x=\d y=0$ and $f(0)=1$.

If we attempt to treat the spaceship as more than a test particle,
we must confront the fact that by construction we have forced
$f=0$ outside the warp bubble. [Consider, for instance, the
explicit form function of equation (\ref{E:form}) in the limit
$r\to\infty$.] This implies that the spacetime geometry is
asymptotically Minkowski space, and in particular the ADM mass
(defined by taking the limit as one moves to spacelike infinity
$i^0$) is zero. That is, the ADM mass of the spaceship and the
warp field generators must be exactly compensated by the ADM mass
due to the stress-energy of the warp-field itself. Viewed in this
light it is now patently obvious that there must be significant
violations of the classical energy conditions (at least in the
original version of the warp-drive spacetime), and the interesting
question becomes ``Where are these energy condition violations
localized?''.

One of our tasks in the current article will be to see if we can
first avoid this exact cancellation of the ADM mass, and second,
to see if we can make qualitative and quantitative statements
concerning the localization and ``total amount'' of energy
condition violations. (A similar attempt at quantification of the
``total amount'' of energy condition violation in traversable
wormholes was recently presented in~\cite{Kar1,Kar2}.)

Consider a spaceship placed within the Alcubierre warp bubble. The
expansion of the volume elements, $\theta=U^{\mu}{}_{;\mu}$, is given
by $\theta=v\;\left({\partial f}/{\partial z} \right)$. Taking into
account equation (\ref{Alcubierreformfunction}), we have (for
Alcubierre's version of the warp bubble)
\begin{equation}
\theta=v\;\frac{z-z_0}{r}\;\frac{\d f(r)}{\d r}.
\end{equation}
The center of the perturbation corresponds to the spaceship's position
$z_0(t)$. The volume elements are expanding behind the spaceship, and
contracting in front of it.
Appendix \ref{A:alcubierre} contains a full calculation of all the
orthonormal components of the Einstein tensor for the Alcubierre warp
bubble. By using the Einstein field equation, $G_{\mu\nu}=8\pi \;
T_{\mu\nu}$, we can make rather general statements regarding the
nature of the stress energy required to support a warp bubble.

The WEC states $T_{\mu\nu} \; V^{\mu} \; V^{\nu}\geq0$, in which
$V^{\mu}$ is \emph{any} timelike vector and $T_{\mu\nu}$ is the
stress-energy tensor.  Its physical interpretation is that the
local energy density is positive.  By continuity it implies the
NEC. In particular the WEC implies that  $T_{\mu\nu} \; U^{\mu} \;
U^{\nu}\geq0$ where $U^\mu$ is the four-velocity of the Eulerian
observers discussed above. The calculations will be simplified
using an orthonormal reference frame. Thus, from the results
tabulated in Appendix \ref{A:alcubierre} we verify that for the
warp drive metric, the WEC is violated, \emph{i.e.},
\begin{equation}
T_{\hat{\mu}\hat{\nu}} \; U^{\hat{\mu}} \; U^{\hat{\nu}}=
-\frac{v^2}{32\pi}\; \left[ \left (\frac {\partial f}{\partial x}
\right )^2 + \left (\frac {\partial f}{\partial y} \right )^2
\right]  <0 \,,
\end{equation}
where $T_{\hat{\mu}\hat{\nu}}$ and $U^{\hat{\mu}}$ are,
respectively, the stress energy tensor and timelike Eulerian
four-velocity given in the orthonormal basis. Taking into account
the Alcubierre form function (\ref{E:form}), we have
\begin{equation}
T_{\hat{\mu}\hat{\nu}} \; U^{\hat{\mu}} \; U^{\hat{\nu}}=
-\frac{1}{32\pi}\frac{v^2 (x^2+y^2)}{r^2} \left[ \frac{\d f}{\d r}
\right]^2<0 \label{WECviolation} \,.
\end{equation}
By considering the Einstein tensor component,
$G_{\hat{t}\hat{t}}$, in an orthonormal basis (details given in
Appendix \ref{A:alcubierre}), and taking into account the Einstein
field equation, we verify that the energy density of the warp
drive spacetime is given by
\begin{equation}
T^{tt} = T^{\mu\nu} \; U_\mu \; U_\nu = T^{\hat t\hat t} =
T_{\hat{t}\hat{t}} = T_{\hat{\mu}\hat{\nu}} \; U^{\hat{\mu}} \;
U^{\hat{\nu}},
\end{equation}
and so equals the energy density measured by the Eulerian
observers, that is, equation (\ref{WECviolation}). It is easy to
verify that the energy density is distributed in a toroidal region
around the $z$-axis, in the direction of travel of the warp
bubble~\cite{PfenningF}.  It is perhaps instructive to point out
that the energy density for this class of spacetimes is nowhere
positive. That the total ADM mass can nevertheless be zero is due
to the intrinsic nonlinearity of the Einstein equations.

We can (in analogy with the definitions in~\cite{Kar1,Kar2}) quantify
the ``total amount'' of energy condition violating matter in the warp
bubble by defining
\begin{equation}
M_\mathrm{warp} = \int \rho_\mathrm{warp} \; \d^3 x =
\int T_{\mu\nu} \; U^{\mu}\; U^{\nu}  \; \d^3 x =
- {v^2\over32\pi} \int \frac{x^2+y^2}{r^2}
\left[ \frac{\d f}{\d r} \right]^2 \; r^2 \; \d r \; \d^2\Omega
=
-{v^2\over12}  \int \left[ \frac{\d f}{\d r} \right]^2 \; r^2 \; \d r.
\end{equation}
This is emphatically not the total mass of the spacetime, but it
characterizes how much (negative) energy one needs to localize in the
walls of the warp bubble. For the specific shape function
(\ref{E:form}) we can estimate
\begin{equation}
M_\mathrm{warp} \approx - v^2 \; R^2 \; \sigma.
\end{equation}
(The integral can be done exactly, but the exact result in terms
of {\sf polylog} functions is unhelpful.) Note that the energy
requirements for the warp bubble scale quadratically with bubble
velocity, quadratically with bubble size, and inversely as the
thickness of the bubble wall.

The NEC states that $T_{\mu\nu} \, k^{\mu} \, k^{\nu}\geq0$, where
$k^{\mu}$ is \emph{any} arbitrary null vector and $T_{\mu\nu}$ is
the stress-energy tensor. Taking into account the Einstein tensor
components presented in Appendix \ref{A:alcubierre}, and the
Einstein field equation, $G_{\mu\nu}=8\pi\; T_{\mu\nu}$, the NEC
for a null vector oriented along the $\pm \hat z$ directions takes
the following form
\begin{equation}
T_{\mu\nu} \; k^{\mu} \; k^{\nu}=
-\frac{v^2}{8\pi}\,
\left[
\left (\frac{\partial f}{\partial x} \right )^2
+
\left (\frac {\partial f}{\partial y} \right )^2
\right]
\pm
\frac{v}{8\pi}\left(
\frac {\partial^{2}f}{\partial {x}^{2}}
+
 \frac {\partial^{2}f}{\partial {y}^{2}}
\right)  \,.
\end{equation}
In particular if we average over the $\pm \hat z$ directions we have
\begin{equation}
{1\over2} \left\{
T_{\mu\nu} \; k^{\mu}_{+\hat z} \; k^{\nu}_{+\hat z} +
T_{\mu\nu} \; k^{\mu}_{-\hat z} \; k^{\nu}_{-\hat z}
\right\}
=
-\frac{v^2}{8\pi}\,
\left[
\left (\frac{\partial f}{\partial x} \right )^2
+
\left (\frac {\partial f}{\partial y} \right )^2
\right],
\end{equation}
which is manifestly negative, and so the NEC is violated for all $v$.
Furthermore, note that even if we do not average, the coefficient of
the term linear in $v$ must be nonzero \emph{somewhere} in the
spacetime. Then at low velocities this term will dominate and at low
velocities the un-averaged NEC will be violated in either the $+\hat
z$ or $-\hat z$ directions.

To be a little more specific about how and where the NEC is violated
consider the Alcubierre form function. We have
\begin{equation}\label{NEC:Alcubierre-form}
T_{\mu\nu} \; k^{\mu}_{\pm\hat z} \; k^{\nu}_{\pm\hat z}=
-\frac{1}{8\pi}
\frac{v^2(x^2+y^2)}{r^2}
\left(
\frac{\d f}{\d r}
\right)^2
\pm
\frac{v}{8\pi}\left[
\frac{x^2+y^2+2(z-z_0(t))^2}{r^3} \;
\frac{\d f}{\d r}+\frac{x^2+y^2}{r^2}\,
\frac{\d^2 f}{\d r^2}\right] \,.
\end{equation}
The first term is manifestly negative everywhere throughout the
space. As $f$ decreases monotonically from the center of the warp
bubble, where it takes the value of $f=1$, to the exterior of the
bubble, with $f\approx 0$, we verify that ${\d f}/{\d r}$ is
negative in this domain. The term ${\d ^2 f}/{\d r^2}$ is also
negative in this region, as $f$ attains its maximum in the
interior of the bubble wall. Thus, the term in square brackets
unavoidably assumes a negative value in this range, resulting in
the violation of the NEC.

For a null vector oriented perpendicular to the direction of motion
(for definiteness take $\hat k = \pm \hat x$) the NEC takes the
following form
\begin{equation}
T_{\mu\nu} \; k^{\mu}_{\pm\hat x} \; k^{\nu}_{\pm\hat x}=
-\frac{v^2}{8\pi}\,
\left[
{1\over2} \left (\frac{\partial f}{\partial y} \right )^2
+ \left(\frac {\partial f}{\partial z} \right )^2
- (1-f) {\partial^2 f\over \partial z^2}
\right]
\mp
\frac{v}{8\pi}\left(
\frac{\partial^{2}f}{\partial {x}\partial z}
\right)  \,.
\end{equation}
Again, note that the coefficient of the term linear in $v$ must be
nonzero \emph{somewhere} in the spacetime. Then at low velocities this
term will dominate, and at low velocities the NEC will be violated in
one or other of the transverse directions.
Upon considering the specific form of the spherically symmetric
Alcubierre form function, we have
\begin{eqnarray}
T_{\mu\nu} \; k^{\mu}_{\pm\hat x} \; k^{\nu}_{\pm\hat x}&=&
-\frac{v^2}{8\pi}
\left[
\frac{y^2+2(z-z_0(t))^2}{2r^2} \left( \frac{\d f}{\d r} \right)^2
-(1-f)
\left(\frac{x^2+y^2}{r^3} \;\frac{\d f}{\d r}
+\frac{(z-z_0(t))^2}{r^2} \; \frac{\d^2f}{\d r^2} \right)
\right]
      \nonumber  \\
&&\mp
\frac{v}{8\pi}\frac{x\,(z-z_0(t))}{r^2}
\left(
\frac{\d^2 f}{\d r^2}-\frac{1}{r}\frac{\d f}{\d r}
\right)
\,.
\end{eqnarray}
Again, the message to take from this is that localized NEC violations
are ubiquitous and persist to arbitrarily low warp bubble velocities.

Using the ``volume integral quantifier'' (as defined
in~\cite{Kar1,Kar2}), we may estimate the ``total amount'' of
averaged null energy condition violating matter in this spacetime,
given by
\begin{eqnarray}
\int T_{\mu\nu} \;k^{\mu}_{\pm\hat z} \; k^{\nu}_{\pm\hat z}  \; \d^3 x \approx
\int T_{\mu\nu} \;k^{\mu}_{\pm\hat x}  \; k^{\nu}_{\pm\hat x}  \; \d^3 x  \approx
- v^2\; R^2 \; \sigma \approx M_{\mathrm{warp}}\,.
\end{eqnarray}
The key things to note here are that the net volume integral of the $O(v)$
term is zero, and that the net volume average of the NEC violations is
approximately the same as the net volume average of the WEC violations,
which are $O(v^2)$.

\subsection{Nat\'{a}rio warp drive}

The alternative version of the warp bubble, due to Jos\'{e}
Nat\'{a}rio demonstrates that the contraction/expansion referred
to above is not always a feature of the warp drive~\cite{Natario}.
In his construction a compression in the radial direction is
exactly balanced by an expansion in the perpendicular direction,
so that the shift vector is divergence-free $\nabla\cdot\beta=0$.
To verify this consider spherical coordinates $(r,\varphi,\phi)$
in the Euclidean 3-space (see \cite{Natario} for details). The
diagonal components of the extrinsic curvature are given by
\begin{eqnarray}
K_{rr}=-2vf'\cos\varphi
  \qquad   {\rm and}   \qquad
K_{\varphi\varphi}=K_{\phi \phi}=vf'\cos\varphi
\label{extrin22}\,,
\end{eqnarray}
respectively, where the prime denotes a derivative with respect to
$r$. Taking into account the definition of the expansion of the
volume element given by $\theta=K^i{}_{i}\,$, from equations
(\ref{extrin22}), we simply have
$\theta=K_{rr}+K_{\varphi\varphi}+K_{\phi \phi}=0$. This analysis
provides a certain insight into the geometry of the spacetime. For
instance, consider the front of the warp bubble, with $\cos\varphi
>0$, such that $K_{rr}<0$ (see \cite{Natario} for details). This
indicates a compression in the radial direction, which is
compensated by an expansion, $K_{\varphi\varphi}+K_{\phi
\phi}=-K_{rr}$, in the perpendicular direction. Thus, this warp
drive spacetime can be thought of as a warp bubble that pushes
space aside and  thereby ``slides'' through space.

We now define $\beta$ in the interior of the warp bubble to be
$v$, and assume the radius of the warp bubble to be $R$.
Furthermore we take $\beta$ to fall to zero over a distance of
order $1/\sigma$. Appendix \ref{A:natario} contains a full
calculation of all the orthonormal components of the Einstein
tensor for the Nat\'{a}rio warp bubble, useful for analysis of
energy condition violations in this class of spacetimes.

For the WEC we use the results of Appendix \ref{A:natario} to write
\begin{equation}
T_{\mu\nu} \; U^{\mu} \; U^{\nu}= {1\over8\pi} G_{\hat t\hat t}
= -  {1\over16\pi}  {\rm Tr}({\bm K}^2).
\end{equation}
This is manifestly negative and so the WEC is violated everywhere
throughout the spacetime.

We can again quantify the ``total amount'' of energy condition
violating matter in the warp bubble by defining
\begin{equation}
M_\mathrm{warp} = \int \rho_\mathrm{warp} \; \d^3 x
=
\int T_{\mu\nu} \; U^{\mu}\; U^{\nu}  \; \d^3 x
=
- {1\over16\pi} \int   {\rm Tr}({\bm K}^2) \;  \d^3 x
=
- {1\over16\pi} \int   \beta_{(i,j)} \;  \beta_{(i,j)} \;  \d^3 x.
\end{equation}
Now $\beta$ in the interior of the warp bubble is by definition $v$,
while the size of the warp bubble is taken to be $R$. Furthermore
gradients of $\beta$ in the bubble walls are of order $v \sigma$, and
the thickness of the bubble walls is of order $1/R$. So we can again
estimate (as for the Alcubierre warp bubble)
\begin{equation}
M_\mathrm{warp} \approx - v^2 \; R^2 \; \sigma.
\end{equation}

With a bit more work we can verify that the NEC is violated throughout
the spacetime. Consider a null vector pointing in the direction $\hat
n$, so that $k^\mu = (1, \hat n^i)$. Then
\begin{equation}
T_{\mu\nu} \; k^{\mu} \; k^{\nu}= {1\over8\pi}
\left[ G_{\hat t\hat t} + 2 \hat n^\ii \; G_{\hat t \ii}
+ \hat n^\ii \hat n^\jj  \; G_{\ii\jj} \right]
\end{equation}
If the NEC is to hold, then this must be positive in both the $\hat n$
and $-\hat n$ directions, so that we must have
\begin{equation}
{1\over2} \left\{
T_{\mu\nu} \; k^{\mu}_{+\hat n} \; k^{\nu}_{+\hat n}
+
T_{\mu\nu} \; k^{\mu}_{-\hat n} \; k^{\nu}_{-\hat n}
\right\}
=
{1\over8\pi}
\left[ G_{\hat t\hat t} + \hat n^i \hat n^j \; G_{\ii\jj} \right] \geq 0
\end{equation}
But now averaging over the $x$, $y$, and $z$ directions we see that
the NEC requires
\begin{equation}
T_{\hat t\hat t} + \sum_\ii T_{\ii\ii} =
{1\over8\pi} [ G_{\hat t\hat t} + \sum_\ii G_{\ii\ii} ] \geq 0.
\end{equation}
But from Appendix \ref{A:natario} we have
\begin{equation}
 G_{\hat t\hat t} + \sum_\ii G_{\ii\ii} =  -2 {\rm Tr}({\bm K}^2)
\end{equation}
which is manifestly negative, and so the NEC is violated everywhere.
Note that ${\bf K}$ is $O(v)$ and so we again see that the NEC
violations persist to arbitrarily low warp bubble velocity.

\section{Linearized gravity applied to the warp drive}

Our goal now  is to try to build a more realistic model of a warp
drive spacetime where the warp bubble is interacting with a finite
mass spaceship. To do so we first consider the linearized theory
\cite{Schutz,Misner,Wald} applied to warp drive spacetimes, for
non-relativistic velocities, $v\ll 1$. A brief summary of
linearized gravity is presented in Appendix \ref{A:linear}.  In
linearized theory, the spacetime metric is given by $\d
s^2=\left(\eta_{\mu\nu} + h_{\mu\nu}\right)\, \d x^\mu \, \d
x^\nu$, with $h_{\mu\nu}\ll 1$ and $\eta_{\mu\nu}={\rm
diag}(-1,1,1,1)$. Taking into account equation
(\ref{Cartesianwarpmetric}), it is an exact statement that
$h_{\mu\nu}$ has the following matrix elements
\begin{equation}
(h_{\mu\nu})=\left[
\begin{array}{cccc}
v^2f^2&0&0&-vf \\
0&0&0&0 \\
0&0&0&0 \\
-vf&0&0&0
\end{array}
\right] \,.
\end{equation}
Now the results deduced from applying linearized theory are only
accurate to first order in $v$. This is equivalent to neglecting the
$h_{00}=v^2f^2$ term in the matrix $\left(h_{\mu\nu}\right)$,
retaining only the first order terms in $v$. Thus, we have the
following approximation
\begin{equation}
(h_{\mu\nu})=\left[
\begin{array}{cccc}
0&0&0&-vf \\
0&0&0&0 \\
0&0&0&0 \\
-vf&0&0&0
\end{array}
\right] \label{linearperturbation}\,.
\end{equation}
The trace of $h_{\mu\nu}$ is identically null, \emph{i.e.},
$h=h^{\mu}{}_{\mu}=0$. Therefore, the trace reverse of $h_{\mu\nu}$,
defined in equation (\ref{tracereverse}), is given by
$\overline{h}_{\mu\nu}=h_{\mu\nu}$, i.e., equation
(\ref{linearperturbation}) itself.

The following relation will be useful in determining the
linearized Einstein tensor,
\begin{equation}
\frac{\partial^2 \overline{h}_{03}}
{\partial x^\alpha \partial x^\beta}=
-v \; \frac{\partial^2 f}{\partial x^\alpha\partial x^\beta}
\label{usefulrel}  \,.
\end{equation}
where we consider the bubble velocity parameter to be constant,
$v=\hbox{constant}$. For simplicity we shall use the original form
function given by Alcubierre, equation (\ref{Alcubierreformfunction}),
so that equation (\ref{usefulrel}) takes the form
\begin{equation}
\frac{\partial^2 \overline{h}_{03}}
{\partial x^\alpha \partial x^\beta}=
-v \left(
\frac{\partial r}{\partial x^\alpha}
\;
\frac{\partial r}{\partial x^\beta}
\;
\frac{\d^2 f}{\d r^2}+
\frac{\partial^2 r}{\partial x^\alpha\partial x^\beta}
\;
\frac{\d f}{\d r}\right)
\label{usefulrelation}\,.
\end{equation}

\subsection{The weak energy condition (WEC)}

The linearized theory applied to Alcubierre's warp drive, for
non-relativistic velocities ($v\ll 1$), is an immediate
application of the linearized Einstein tensor, equation
(\ref{linearEinstein}), and there is no need to impose the Lorenz
gauge~\cite{Lorenz}. The interest of this exercise lies in the
application of the WEC. In linearized theory the $4$-velocity can
be approximated by $U^{\mu}=(1,0,0,0)$, therefore the WEC reduces
to
\begin{equation}
T_{\mu\nu} \; U^{\mu} \; U^{\nu}=T_{00}=\frac{1}{8\pi}\; G_{00}
\label{linearwarpWEC} \,,
\end{equation}
and from equation (\ref{linearEinstein}), we have
\begin{equation}
G_{00}=-\frac{1}{2}\left(
\overline{h}_{00,\mu}{}^{\mu}-
\overline{h}_{\mu\nu,}{}^{\mu\nu}-
2 \; \overline{h}_{0 \mu,0}{}^{\mu}\right)
\label{Einsteincomp00}  \,.
\end{equation}
The respective terms are given by
\begin{eqnarray}
\overline{h}_{00,\mu}{}^{\mu}&=&0
\label{trace1}  \,,
\\
\overline{h}_{\mu\nu,}{}^{\mu\nu}&=&-2 \overline{h}_{03,03}
\label{trace2} \,,
\\
\overline{h}_{0 \mu,0}{}^{\mu}&=& \overline{h}_{03,03}
\label{trace3} \,.
\end{eqnarray}

In general, if second time derivatives appear, they can be neglected
because ${\partial}/{\partial t}$ is of the same order as $v \;
{\partial}/{\partial z}$, so that
$\partial_{\mu}\partial^{\mu}=\nabla^2+O(v^2 \;\nabla^2))$. In fact,
it is possible to prove that the Einstein tensor components
$G_{00}$ and $G_{0i}$ do not contain second time derivatives of any
generic $\overline{h}_{\mu\nu}$~\cite{Schutz}.  That is, only the six
equations $G_{ij}=8\pi T_{ij}$, are true dynamical equations.  In
contrast, the equations $G_{0\mu}=8\pi T_{0\mu}$ are called {\it
constraint equations} because they are relations among the initial
data for the other six equations; which prevent one from freely
choosing the initial data.

Substituting equations (\ref{trace1})--(\ref{trace3}) into equation
(\ref{Einsteincomp00}), we have
\begin{equation}\label{G00}
G_{00}= O(v^2) \,.
\end{equation}
Thus, equation (\ref{linearwarpWEC}) is given by
\begin{equation}
T_{\mu\nu} \; U^{\mu} \; U^{\nu}=T_{00}=O(v^2)
\label{zerolinearwarpWEC} \,,
\end{equation}
and the WEC is identically ``saturated''.  Although in this
approximation the WEC is not violated, it is on the verge of being
so (to first order in $v$). This is compatible with the ``exact''
non-perturbative calculation previously performed.

\subsection{Negative energy density in boosted inertial frames}

Despite the fact that the observers, with $U^{\mu}=(1,0,0,0)$,
measure zero energy density [more precisely $O(v^2)$], it can be
shown that observers which move with any other arbitrary velocity,
$\tilde\beta$, along the positive $z$ axis measure a negative
energy density [at $O(v)$]. That is, $T_{\hat{0}\hat{0}}<0$. The
$\tilde\beta$ occurring here is completely independent of the
shift vector $\beta(x,y,z-z_0(t))$, and is also completely
independent of the warp bubble velocity $v$. Consider a Lorentz
transformation, $x^{\hat{\mu}} = \Lambda^{\hat{\mu}}{}_{\nu} \;
x^{\nu}$, with $\Lambda^{\mu}{}_{\hat{\alpha}} \;
\Lambda^{\hat{\alpha}}{}_{\nu} = \delta^{\mu}{}_{\nu}$ and
$\Lambda^{\mu}{}_{\hat{\nu}}$ defined as
\begin{equation}
(\Lambda^{\mu}{}_{\hat{\nu}})=\left[
\begin{array}{cccc}
\gamma&0&0&\gamma\tilde\beta \\
0&1&0&0 \\
0&0&1&0 \\
\gamma\tilde\beta&0&0&\gamma
\end{array}
\right]   \label{Lorentzmatrix}\,,
\end{equation}
with $\gamma=(1-\tilde\beta^2)^{-1/2}$. The energy density
measured by these observers is given by $T_{\hat{0}\hat{0}} =
\Lambda^{\mu}{}_{\hat{0}}\; \Lambda^{\nu}{}_{\hat{0}}\;
T_{\mu\nu}$. That is:
\begin{eqnarray}
T_{\hat{0}\hat{0}}
&=&
\gamma^{2}T_{00}+2\gamma^{2} \tilde\beta T_{03}+\gamma^{2}\tilde\beta^{2}T_{33}
\\
\nonumber
&=&
\gamma^{2}\tilde\beta\left(2T_{03}+\tilde\beta T_{33}\right)
\\
&=&
\frac{\gamma^{2}\tilde\beta}{8\pi}\left(2G_{03}+\tilde\beta G_{33}\right)
\label{observerenergydensity}\,.
\end{eqnarray}
taking into account $T_{00}=0$, due to equation
(\ref{zerolinearwarpWEC}).  The respective Einstein tensor components,
$G_{03}$ and $G_{33}$, are given by
\begin{eqnarray}
G_{03}&=&-\frac{1}{2}\left(\overline{h}_{03,11}
+\overline{h}_{03,22}\right)   + O(v^2)
\label{Einstein03}  \,,\\
G_{33}&=& O(v^2)
\label{Einstein33}  \,.
\end{eqnarray}

Finally, substituting these into equation (\ref{observerenergydensity}), to a
first-order approximation in terms of $v$, we have
\begin{equation}
T_{\hat{0}\hat{0}}=-\frac{\gamma^{2}\tilde\beta}{8\pi}\left(\overline{h}_{03,11}+
\overline{h}_{03,22}\right) + O(v^2)
\label{observerenergy2}\,.
\end{equation}

Taking into account equation (\ref{usefulrel}), we have
\begin{equation}
T_{\hat{0}\hat{0}}=\frac{\gamma^{2}\tilde\beta
v}{8\pi}\left(\frac{\partial^2 f}{\partial x^2}+\frac{\partial^2
f}{\partial y^2}\right)  + O(v^2)
\label{negenergy2} \,.
\end{equation}

Applying equation (\ref{usefulrelation}), we have the following
relations
\begin{eqnarray}
\overline{h}_{03,11}&=&-v \left[\frac{x^2}{r^2} \frac{\d^2
f}{\d r^2}+\left(\frac{1}{r}-\frac{x^2}{r^3}
\right)\frac{\d f}{\d r}\right] \,,\\
\overline{h}_{03,22}&=&-v \left[\frac{y^2}{r^2} \frac{\d^2
f}{\d r^2}+\left(\frac{1}{r}-\frac{y^2}{r^3} \right)
\frac{\d f}{\d r}\right]\,,
\end{eqnarray}
and substituting in equation (\ref{observerenergy2}) we therefore have
\begin{equation}
T_{\hat{0}\hat{0}}=
\frac{\gamma^{2}\tilde\beta v}{8\pi}
\left[
\left(\frac{x^2+y^2}{r^2}\right)
\;
\frac{\d^2 f}{\d r^2}
+
\left(\frac{x^2+y^2+2(z-z_0(t))^2}{r^3}\right)
\;
\frac{\d f}{\d r}
\right]  + O(v^2)
\label{negativeenergydensity}\,.
\end{equation}
A number of general features can be extracted from the terms in
square brackets, without specifying an explicit form of $f$. In
particular, $f$ decreases monotonically from its value at $r=0$,
$f=1$, to $f\approx 0$ at $r\geq R$, so that ${\d f}/{\d r}$ is
negative in this domain. The form function attains its maximum in
the interior of the bubble wall, so that ${\d^2 f}/{\d r^2}$ is
also negative in this region. Therefore there is a range of $r$ in
the immediate interior neighbourhood of the bubble wall that
necessarily provides negative energy density, as seen by the
observers considered above. Again we find that WEC violations
persist to arbitrarily low warp bubble velocities.

\subsection{The null energy condition (NEC)}

The NEC states that $T_{\mu\nu} \, k^{\mu} \, k^{\nu}\geq0$, where
$k^{\mu}$ is a null vector. Considering
$k^{\hat\mu}=(1,0,0,\pm1)$, we have
\begin{eqnarray}
T_{\mu\nu}k^{\mu}k^{\nu}&=&\frac{1}{8\pi}\left(G_{00} \pm
2G_{03}+G_{33}\right)
            \nonumber     \\
&=&\mp\frac{1}{8 \pi}\left(\overline{h}_{03,11} +
\overline{h}_{03,22}\right) + O(v^2)
            \nonumber      \\
&=&\pm\frac{v}{8 \pi}\left(\frac{\partial^2 f}{\partial
x^2}+\frac{\partial^2 f}{\partial y^2}\right) + O(v^2)
\label{observerlinearNEC}  \,.
\end{eqnarray}
Whatever the value of the bracketed term, as long as it is nonzero
the NEC will be violated (for small enough $v$) in either the $+z$
or $-z$ directions.  This is compatible with the ``exact''
non-perturbative calculation previously performed.

Using the general form function defined by Alcubierre, equation
(\ref{observerlinearNEC}) reduces to
\begin{equation}
T_{\mu\nu}k^{\mu}k^{\nu}= \pm
\frac{v}{8\pi}\left[
\left(\frac{x^2+y^2}{r^2}\right) \;
\frac{\d^2f}{\d r^2}+
\left(\frac{x^2+y^2+2(z-z_0(t))^2}{r^3}\right)\;
\frac{\d f}{\d r}\right] + O(v^2)
\label{linearNEC}
\end{equation}
Equation (\ref{linearNEC}) is proportional to the energy density,
$T_{\hat{0}\hat{0}}$, of equation (\ref{negativeenergydensity}).
We verify that the term in square brackets has a region for which
it is negative, thus also violating the NEC in the immediate
interior vicinity of the bubble wall.

\subsection{Linearized gravity applied to the spaceship}

The weak gravitational field of a static source, in particular of a
spaceship, is given by the following metric
\begin{equation}
\d s^2=-\d t^2+\d x^2+\d y^2+\d z^2
-2\Phi(x,y,z)\left(\d t^2+\d x^2+\d y^2+\d z^2\right)\,.
\label{staticspaceshipmetric}
\end{equation}
Applying the linearized theory with $\d
s^2=(\eta_{\mu\nu}+h_{\mu\nu}) \, \d x^\mu \, \d x^\nu$ and
$h_{\mu\nu}\ll 1$, the matrix elements of $h_{\mu\nu}$ are given
by
\begin{equation}
(h_{\mu\nu})=\left[
\begin{array}{cccc}
-2\Phi&0&0&0 \\
0&-2\Phi&0&0 \\
0&0&-2\Phi&0 \\
0&0&0&-2\Phi
\end{array}
\right] \,.
\end{equation}
The trace is given by $h=h^{\mu}{}_{\mu}=-4\Phi$. The elements of
the trace reverse, $\overline{h}_{\mu\nu}$, are the following
\begin{equation}
(\overline{h}_{\mu\nu})=\left[
\begin{array}{cccc}
-4\Phi&0&0&0 \\
0&0&0&0 \\
0&0&0&0 \\
0&0&0&0
\end{array}
\right] \,.
\end{equation}
Applying the linearized Einstein tensor, equation
(\ref{linearEinstein}), to the component, $G_{00}$, we have Poisson's
equation
\begin{equation}\label{Poissoneq}
\nabla ^2\Phi =4\pi \rho  \,,
\end{equation}
in which $\rho$ is the mass density of the spaceship. This case is
comparable to the Newtonian limit, which is valid when the
gravitational fields are too weak to produce velocities near the speed
of light, that is, $|\Phi| \ll 1$ and $|v| \ll 1$. For such cases
general relativity makes the same predictions as Newtonian
gravity. For fields that change due to the movement of the source, for
instance along the $z$-axis with a velocity of $v$, we verify that
$\partial /\partial t$ is of the same order as $v\,\partial /\partial
z$. It is easy to verify that the energy fluxes or the momentum
densities, $T_{0i}$ are proportional to first order in $v$, and the
stresses, $T_{ij}$, to second order. Therefore the stress-energy
tensor $T_{\mu\nu}$ obeys the inequalities $|T_{00}|\gg
|T_{0i}|\gg |T_{ij}|$.

\subsection{Spaceship immersed in the warp bubble}

Consider now a spaceship in the interior of an Alcubierre warp bubble,
which is moving along the positive $z$ axis with a non-relativistic
constant velocity.  That is, $v\ll 1$. The metric is given by
\begin{eqnarray}
\lefteqn{
\d s^2=-\d t^2+\d x^2+\d y^2+\left[\d z-v\;f(x,y,z-vt)\,\d t \right]^2
}
\nonumber   \\
&&-2\Phi(x,y,z-vt)\,
\left[\d t^2+\d x^2+\d y^2+(\d z-v\;f(x,y,z-vt)\,\d t)^2\right]
\label{warpspaceshipmetric}  \,.
\end{eqnarray}
If $\Phi =0$, the metric (\ref{warpspaceshipmetric}) reduces to the
warp drive spacetime of equation (\ref{Cartesianwarpmetric}). If
$v=0$, we have the metric representing the gravitational field of a
static source, equation (\ref{staticspaceshipmetric}).

Applying the transformation, $z'=z-vt$, the metric
(\ref{warpspaceshipmetric}) takes the form
\begin{eqnarray}
\lefteqn{ \d s^2=-\d t^2+\d x^2+\d y^2
+\left[\d z'-v\;\{f(x,y,z')-1\}\,\d t
\right]^2 }
\nonumber   \\
&&-2\Phi(x,y,z')\,
\left[\d t^2+\d x^2+\d y^2+(\d z'-v\;\{f(x,y,z')-1\}\,\d t)^2\right]
\label{warpspaceshipmetric2}  \,.
\end{eqnarray}
Note that the metric now looks ``static''. Considering an observer
in the interior of the bubble, co-moving with the spaceship, the
metric (\ref{warpspaceshipmetric2}), with $f=1$, reduces to
\begin{equation}
\label{E:inside}
\d s^2\to -\d t^2+\d x^2+\d y^2+\d z'^2-2\Phi(x,y,z')
\left[\d t^2+\d x^2+\d y^2+\d z'^2 \right] \,,
\end{equation}
so that inside the bubble we simply have the weak gravitational field
of a static source, that of the spaceship.

Outside the warp bubble, where $f=0$, the metric
(\ref{warpspaceshipmetric2}) reduces to
\begin{eqnarray}
\d s^2\to -\d t^2+\d x^2+\d y^2+\left[\d z'+v\,\d t \right]^2
-2\Phi(x,y,z')\,
\left[\d t^2+\d x^2+\d y^2+(\d z'+v\,\d t)^2\right]
\label{E:outside}  \,.
\end{eqnarray}

\subsubsection{First order approximation}

Applying the linearized theory, keeping terms linear in $v$ and
$\Phi$ but neglecting all superior order terms so that the
approximation of equation (\ref{linearRicci}) is valid, the matrix
elements, $h_{\mu\nu}$, of the metric (\ref{warpspaceshipmetric})
are given by the following approximation
\begin{equation}
(h_{\mu\nu})=\left[
\begin{array}{cccc}
-2\Phi&0&0&-vf \\
0&-2\Phi&0&0 \\
0&0&-2\Phi&0 \\
-vf&0&0&-2\Phi
\end{array}
\right]   \,.
\end{equation}
The trace of $h_{\mu\nu}$ is given by,
$h=h^{\mu}_{\;\;\mu}=-4\Phi$, and the trace-reversed elements,
$\overline{h}_{\mu\nu}$, take the following form
\begin{equation}
(\overline{h}_{\mu\nu})=\left[
\begin{array}{cccc}
-4\Phi&0&0&-vf \\
0&0&0&0 \\
0&0&0&0 \\
-vf&0&0&0
\end{array}
\right]   \,.
\end{equation}

Therefore, from the WEC, $T_{\mu\nu}U^{\mu}U^{\nu}=T_{00}$, the
trace-reversed elements that one needs to determine $G_{00}$ [see
equation (\ref{Einsteincomp00})] are the following
\begin{eqnarray}
\overline{h}_{00,\mu}{}^{\mu}&=&-\overline{h}_{00,00}+\overline{h}_{00,11}+
\overline{h}_{00,22}+\overline{h}_{00,33} \label{h1} \,,
\\
\overline{h}_{\mu\nu,}{}^{\mu\nu}&=&\overline{h}_{00,00}-2
\overline{h}_{03,03} \label{h2} \,,
\\
\overline{h}_{0 \mu,0}{}^{\mu}&=&-\overline{h}_{00,00}+
\overline{h}_{03,03} \label{h3} \,.
\end{eqnarray}

Substituting equations (\ref{h1})--(\ref{h3}) into equation
(\ref{Einsteincomp00}), we have
\begin{equation}\label{G00linear}
G_{00}=2 \, \nabla ^2\Phi  + O(v^2, v \Phi, \Phi^2) \,.
\end{equation}
Thus, taking into account Poisson's equation, we verify that the
WEC is given by
\begin{equation}
T_{\mu\nu}U^{\mu}U^{\nu}=\rho  + O(v^2, v \Phi, \Phi^2) \,,
\end{equation}
where $\rho$ is now the ordinary energy density of the spaceship which
is manifestly positive.  In linearized theory, the total ADM mass of
the space-time simply reduces to the mass of the space-ship, i.e.,
\begin{equation}
M_\mathrm{ADM}=\int T_{00} \, \d ^3 x=\int \rho \; \d ^3 x  +
O(v^2, v \Phi, \Phi^2) = M_\mathrm{ship} + O(v^2, v \Phi,
\Phi^2)\,.
\end{equation}
The good news for the warp drive aficionados is that the dominant
term is manifestly positive.

The NEC, with $k^{\mu}\equiv (1,0,0,\pm 1)$, takes the form
\begin{eqnarray}
T_{\mu\nu}k^{\mu}k^{\nu}=\frac{1}{8\pi}\left(G_{00}\pm
2G_{03}+G_{33}\right) \label{observerlinearNEC2}  \,.
\end{eqnarray}
The Einstein tensor components, $G_{03}$ and $G_{33}$, are given by
\begin{eqnarray}
G_{03}&=&-\frac{1}{2}\left(\overline{h}_{03,\mu}
^{\;\;\;\;\;\;\;\mu}-\overline{h}_{0\mu,3}^{\;\;\;\;\;\;\;\mu}-
\overline{h}_{3 \mu,0}^{\;\;\;\;\;\;\;\mu}\right)
      \label{2Einstein03}  \,,\\
G_{33}&=&-\frac{1}{2}\left(\overline{h}_{\mu\nu,}
^{\;\;\;\;\;\;\mu\nu}-2 \overline{h}_{3
\mu,3}^{\;\;\;\;\;\;\;\mu}\right) \label{2Einstein33}  \,,
\end{eqnarray}
with the respective trace-reversed terms
\begin{eqnarray}
\overline{h}_{03,\mu}
^{\;\;\;\;\;\;\;\mu}&=&-\overline{h}_{03,00}+\overline{h}_{03,11}+
\overline{h}_{03,22}+\overline{h}_{03,33}      \\
\overline{h}_{0\mu,3}
^{\;\;\;\;\;\;\;\mu}&=&-\overline{h}_{00,30}+\overline{h}_{03,33}
    \\
\overline{h}_{3\mu,0} ^{\;\;\;\;\;\;\;\mu}&=&-\overline{h}_{30,00}
    \\
\overline{h}_{3\mu,3} ^{\;\;\;\;\;\;\;\mu}&=&-\overline{h}_{30,30}
\end{eqnarray}
and the term $\overline{h}_{\mu\nu,}{}^{\mu\nu}$ is given by equation
(\ref{h2}). Substituting these into equations
(\ref{2Einstein03})--(\ref{2Einstein33}), we have
\begin{eqnarray}
G_{03}&=&\frac{v}{2} \left(\frac{\partial^2 f}{\partial
x^2}+\frac{\partial^2 f}{\partial y^2} \right)
+ O(v^2, v \Phi, \Phi^2)  \,,  \\
G_{33}&=&O(v^2, v \Phi, \Phi^2) \,.
\end{eqnarray}
to first order in $v$ and $\Phi$, and neglecting the crossed terms $v
\, \Phi$.

The null energy condition is thus given by
\begin{equation}
T_{\mu\nu} k^{\mu} k^{\nu}=\rho
\pm\frac{v}{8\pi}\left(\frac{\partial^2 f}{\partial
x^2}+\frac{\partial^2 f}{\partial y^2}\right) + O(v^2, v \Phi,
\Phi^2) \,.
\end{equation}
From this, one can deduce the existence of localized NEC
violations even in the presence of a finite mass spaceship, and
can also make deductions about the net volume-averaged NEC
violations. First, note that for reasons of structural integrity
one wants the spaceship itself to lie well inside the warp bubble,
and not overlap with the walls of the warp bubble.  But this means
that the region where $\rho\neq0$ does not overlap with the region
where the $O(v)$ contribution due to the warp field is non-zero.
So regardless of how massive the spaceship itself is, there will
be regions in the wall of the warp bubble where localized
violations of NEC certainly occur.  If we now look at the volume
integral of the NEC, the $O(v)$ contributions integrate to zero
and we have
\begin{eqnarray}
\int T_{\mu\nu} \;k^{\mu}_{\pm\hat z} \; k^{\nu}_{\pm\hat z}  \; \d^3 x =
\int \rho \; \d ^3 x  + O(v^2,v\Phi,\Phi^2) = M_\mathrm{ship} + O(v^2,v\Phi,\Phi^2)\,.
\end{eqnarray}

\medskip

Consider a similar analysis for a null vector oriented perpendicular
to the direction of motion, for instance, along the $x-$axis, so that
$k^\mu=(1,\pm 1,0,0)$. The NEC then takes the form
\begin{equation}
T_{\mu\nu}k^{\mu}k^{\nu}=\frac{1}{8\pi}\left(G_{00}\pm
2G_{01}+G_{11}\right) \label{observerlinearNEC:x}  \,.
\end{equation}

Taking into account equation (\ref{linearEinstein}), the
respective Einstein tensor components are given by
\begin{eqnarray}
G_{01}&=&-\frac{v}{2}\,\frac{\partial^2 f}{\partial
x \partial z} +O(v^2, v \Phi, \Phi^2) \,,  \\
G_{11}&=&O(v^2, v \Phi, \Phi^2) \,,
\end{eqnarray}
and $G_{00}$ is given by equation (\ref{G00linear}). Thus,
equation (\ref{observerlinearNEC:x}) takes the following form
\begin{equation}
T_{\mu\nu}k^{\mu}k^{\nu}=\rho \mp \frac{v}{8\pi} \frac{\partial^2
f}{\partial x \partial z}  + O(v^2, v \Phi, \Phi^2) \,,
\end{equation}
and considering the Alcubierre form function, we have
\begin{equation}
T_{\mu\nu}k^{\mu}k^{\nu}=\rho \mp
\frac{v}{8\pi}\frac{x\,(z-z_0(t))}{r^2} \left( \frac{\d^2 f}{\d
r^2}-\frac{1}{r}\frac{\d f}{\d r} \right)  + O(v^2, v \Phi, \Phi^2) \,.
\end{equation}
As for the situation when we considered null vectors aligned with the
direction of motion, for these transverse null vectors we find localized
NEC violations in the walls of the warp bubble. We also find that the
volume integral of the $O(v)$ term is zero and that
\begin{eqnarray}
\int T_{\mu\nu} \;k^{\mu}_{\pm\hat x} \; k^{\nu}_{\pm\hat x}  \; \d^3 x =
\int \rho \; \d ^3 x  + O(v^2,v\Phi,\Phi^2) = M_\mathrm{ship} + O(v^2,v\Phi,\Phi^2)\,.
\end{eqnarray}

The net result of this $O(v)$ calculation is that irrespective of the
mass of the spaceship there will always be localized NEC violations
in the wall of the warp bubble, and these localized NEC violations
persist to arbitrarily low warp velocity. However at $O(v)$ the volume
integral of the NEC violations is zero, and so we must look at higher
order in $v$ if we wish to deduce anything from the consideration of
volume integrals to probe ``net'' violations of the NEC.

\subsubsection{Second order approximation}


Consider the approximation in which we keep the exact $v$
dependence but linearize in the gravitational field of the
spaceship $\Phi$. The components of the Einstein tensor, relevant
to determining the WEC and the NEC are

\begin{eqnarray}
G_{\hat{t}\hat{t}} &=& - \frac{v^2}{4}\left[\left(\frac{\partial
f}{\partial x}\right)^2 + \left(\frac{\partial f}{\partial y}
\right)^2\right] + 2\nabla ^2\Phi +O(\Phi^2)   \,,
\\
G_{\hat{z}\hat{z}} &=& -\frac{3}{4}v^2\left[ \left(\frac{\partial
f}{\partial x}\right)^2+\left(\frac{\partial f}{\partial
y}\right)^2 \right] +O(\Phi^2)   \,,
\\
G_{\hat{t}\hat{z}} &=& \frac{v}{2}\left(\frac{\partial^2
f}{\partial x^2} + \frac{\partial^2 f}{\partial y^2}\right)
+O(\Phi^2)    \,,
\\
G_{\hat{t}\hat{x}} &=& -\frac{1}{2}\,v{\frac
{\partial^{2}\,f}{\partial x \partial z}} +O(\Phi^2) \,,
\\
G_{\hat{x}\hat{x}} &=& v^2 \left[ \frac{1}{4}\,\left ( {\frac
{\partial f}{\partial x}} \right )^{2} - \frac{1}{4}\,\left
({\frac {\partial f}{\partial y}} \right )^{2} - \left({\frac
{\partial f}{\partial z}}\right )^{2} + (1-f){\frac {\partial
^{2}f}{\partial {z}^{2}}}  \right] +O(\Phi^2)  \,.
\end{eqnarray}

The WEC is given by
\begin{eqnarray}
T_{\hat{\mu}\hat{\nu}} \; U^{\hat{\mu}} \; U^{\hat{\nu}}=\rho -
\frac{v^2}{32\pi}\left[\left(\frac{\partial f}{\partial
x}\right)^2 + \left(\frac{\partial f}{\partial y}\right)^2\right]
+O(\Phi^2)
 \,,
\end{eqnarray}
or by taking into account the Alcubierre form function, we have
\begin{equation}
T_{\mu\nu} \; U^{\mu}\; U^{\nu}=\rho -\frac{1}{32\pi}\frac{v^2
(x^2+y^2)}{r^2} \left( \frac{\d f}{\d r} \right)^2 +O(\Phi^2)
 \,.
\end{equation}

Once again, using the ``volume integral quantifier'', we find the
following estimate
\begin{eqnarray}
\int T_{\hat{\mu}\hat{\nu}} \; U^{\hat{\mu}} \; U^{\hat{\nu}} \;
\d^3 x =  M_{\rm ship} -v^2 \;R^2\;\sigma
+ \int O(\Phi^2)  \;\d^3 x\,,
\end{eqnarray}
which we can recast as
\begin{equation}
M_\mathrm{total} = M_\mathrm{ship} + M_\mathrm{warp} + \int
O(\Phi^2) \; \d^3 x\,,
\end{equation}
where $M_\mathrm{total}$ is the net total integrated energy
density, to the order of approximation considered, i.e., keeping
the exact $v$ dependence and linearizing in the gravitational
field $\Phi$.

Now suppose we demand that the volume integral of the WEC at least
be positive, then
\begin{equation}
v^2 \;R^2\;\sigma  \lesssim M_{\rm ship} .
\end{equation}
This equation is effectively the quite reasonable  condition that
the net total energy stored in the warp field be less than the
total mass-energy of the spaceship itself, which places a powerful
constraint on the velocity of the warp bubble. Re-writing this in
terms of the size of the spaceship $R_\mathrm{ship}$ and the
thickness of the warp bubble walls $\Delta = 1/\sigma$, we have
\begin{equation}
v^2 \lesssim {M_{\rm ship}\over R_\mathrm{ship}}\;
{R_\mathrm{ship} \; \Delta\over R^2}.
\end{equation}
For any reasonable spaceship this gives extremely low bounds on
the warp bubble velocity.

\medskip

In a similar manner, the NEC, with $k^\mu=(1,0,0,\pm 1)$, is given by
\begin{eqnarray}
T_{\hat{\mu}\hat{\nu}} \; k^{\hat{\mu}} \; k^{\hat{\nu}} &=& \rho
\pm \frac{v}{8\pi}\left(\frac{\partial^2 f}{\partial x^2} +
\frac{\partial^2 f}{\partial y^2}\right) -
\frac{v^2}{8\pi}\left[\left(\frac{\partial f}{\partial x}\right)^2
+ \left(\frac{\partial f}{\partial y} \right)^2\right] + O(\Phi^2)
\,.
\end{eqnarray}
Considering the ``volume integral quantifier'', we verify that, as
before, the exact solution in terms of polylogarithmic functions
is unhelpful, although we may estimate that
\begin{equation}
\int T_{\hat{\mu}\hat{\nu}} \; k^{\hat{\mu}} \; k^{\hat{\nu}} \;
\d^3 x = M_{\rm ship}  -v^2 \, R^2
\, \sigma   + \int O(\Phi^2) \; \d^3 x\,,
\end{equation}
which is [to order $O(\Phi^2)$] the same integral we encountered
when dealing with the WEC. This volume integrated NEC is now
positive if
\begin{equation}
v^2 \;R^2\;\sigma  \lesssim M_{\rm ship} .
\end{equation}

Finally, considering a null vector oriented perpendicularly to the
direction of motion (for definiteness take $\hat k = \pm \hat x$),
the NEC takes the following form
\begin{equation}
T_{\hat{\mu}\hat{\nu}} \; k^{\hat{\mu}} \; k^{\hat{\nu}}=\rho
-\frac{v^2}{8\pi}\, \left[ {1\over2} \left (\frac{\partial
f}{\partial y} \right )^2 + \left(\frac {\partial f}{\partial z}
\right )^2 - (1-f) {\partial^2 f\over \partial z^2} \right] \mp
\frac{v}{8\pi}\left( \frac{\partial^{2}f}{\partial {x}\partial z}
\right) + O(\Phi^2) \,.
\end{equation}

Once again, evaluating the ``volume integral quantifier'', we have
\begin{equation}
\int T_{\hat{\mu}\hat{\nu}} \; k^{\hat{\mu}} \; k^{\hat{\nu}} \;
\d^3 x =M_{\rm ship} -{v^2\over4}  \int \left( \frac{\d f}{\d r}
\right)^2 \; r^2 \; \d r  + \frac{v^2}{6} \int
(1-f)\,\left(2r\frac{\d f}{\d r}+r^2\frac{\d^2f}{\d r^2} \right)
\, \d r    + \int O(\Phi^2) \d^3 x \,,
\end{equation}
which, as before, may be estimated as
\begin{equation}
\int T_{\hat{\mu}\hat{\nu}} \; k^{\hat{\mu}} \; k^{\hat{\nu}} \;
\d^3 x \approx  M_{\rm ship} -v^2 R^2\, \sigma     + \int O(\Phi^2) \; \d^3 x\,.
\end{equation}
If we do not want the total NEC violations in the warp field to exceed
the mass of the spaceship itself we must again demand
\begin{equation}
v^2 \;R^2\;\sigma  \lesssim M_{\rm ship} ,
\end{equation}
which places a strong constraint on the velocity of the warp bubble.

\section{Summary and Discussion}

In this article we have seen how the warp drive spacetimes (in
particular, the Alcubierre and Nat\'{a}rio warp drives) can be
used as gedanken-experiments to probe the foundations of general
relativity. Though they are useful toy models for theoretical
investigations, as potential technology they are greatly lacking.
We have verified that the non-perturbative exact solutions of the
warp drive spacetimes necessarily violate the classical energy
conditions, and continue to do so for arbitrarily low warp bubble
velocity --- thus the energy condition violations in this class of
spacetimes is generic to the form of the geometry under
consideration and is not simply a side-effect of the
``superluminal''  properties.

Furthermore, by taking into account the notion of the ``volume
integral quantifier'', we have also verified that the ``total
amount'' of energy condition violating matter in the warp bubble
is negative. Using linearized theory, we have built a more
realistic model of the warp drive spacetime where the warp bubble
interacts with a finite mass spaceship. We have tested and
quantified  the energy conditions to first and second order of the
warp bubble velocity. By doing so we have been able to safely
ignore the causality problems associated with ``superluminal''
motion, and so have focussed attention on a previously unremarked
feature of the ``warp drive'' spacetime.  If it is possible to
realise even a weak-field warp drive in nature, such a spacetime
appears to be an example of a ``reaction-less drive''. That is,
the warp bubble moves by interacting with the geometry of
spacetime instead of expending reaction mass, and the spaceship
(which in linearized theory can be treated as a finite mass object
placed within the warp bubble), is simply carried along with it.
We have verified that in this case, the ``total amount'' of energy
condition violating matter (the ``net'' negative energy of the
warp field) must be an appreciable fraction of the positive mass
of the spaceship carried along by the warp bubble. This places an
extremely stringent condition on the warp drive spacetime, namely,
that for all conceivably interesting situations the bubble
velocity should be absurdly low, and it therefore appears unlikely
that, by using this analysis, the warp drive will ever prove to be
technologically useful. Finally, we point out that any attempt at
building up a ``strong-field'' warp drive starting from an
approximately Minkowski spacetime will inevitably have to pass
through a weak-field regime. Since the weak-field warp drives are
already so tightly constrained, the analysis of this article
implies additional difficulties for developing a ``strong field''
warp drive. In particular we wish to sound a strong cautionary
note against over-enthusiastic mis-interpretation of the
technological situation.

\section*{Acknowledgements}

The research of Matt Visser was supported by the Marsden Fund
administered by the Royal Society of New Zealand. Francisco Lobo
acknowledges some helpful discussions with Jos\'{e} Nat\'{a}rio.

\appendix
\section{The Einstein tensor for the Alcubierre warp drive}
\label{A:alcubierre}
\subsection{Generic form function}

The Einstein tensor is given by $G_{\mu\nu} = R_{\mu\nu}-\frac{1}{2}
\; g_{\mu\nu} \; R$. The Einstein tensor components of the warp drive
spacetime, in cartesian coordinates, in an orthonormal basis,
$G_{\hat{\mu}\hat{\nu}}$, with a generic form function,
$f(x,y,z-z_0(t))$, are given by
\begin{eqnarray}
G_{\hat{t}\hat{t}}
&=&
-
\frac{1}{4}\,v^{2}
\left[
\left(\frac {\partial f}{\partial x} \right )^2
+
\left (\frac {\partial f}{\partial y} \right )^2
\right] \,,
\\
G_{\hat{t}\hat{x}}
&=&
-\frac{1}{2}\,v{\frac {\partial^{2}\,f}{\partial x \partial z}}  \,,
\\
G_{\hat{t}\hat{y}}
&=
&-\frac{1}{2}\,v{\frac {\partial^{2}\,f}{\partial y \partial z}}  \,,
\\
G_{\hat{t}\hat{z}}
&=&
\frac{1}{2}\,v\left(
{\frac {\partial ^{2}f}{\partial {x}^{2}}}
+\frac {\partial ^{2}f}{\partial {y}^{2}}
\right)  \,,
\\
G_{\hat{x}\hat{x}}
&=&
v^2 \left[ \frac{1}{4}\,\left (
{\frac {\partial f}{\partial x}} \right )^{2}
-
\frac{1}{4}\,\left
({\frac {\partial f}{\partial y}} \right )^{2}
-
\left({\frac
{\partial f}{\partial z}}\right )^{2}
+
(1-f){\frac {\partial
^{2}f}{\partial {z}^{2}}}  \right] \,,
\\
G_{\hat{x}\hat{y}}
&=&
\frac{1}{2}\,v^2\left ({\frac {\partial f}{\partial y}}\right )
\left(\frac {\partial f}{\partial x}\right)\,,
\\
G_{\hat{x}\hat{z}}
&=&
{v}^ {2} \left[
\left({\frac {\partial f}{\partial z}}\right )
\left(\frac {\partial f}{\partial x}\right)
-
\frac{1}{2}\,(1-f){\frac {\partial ^{2}f}{\partial
x\partial z}}
\right]
 \,,
\\
G_{\hat{y}\hat{z}}
&=&
{v}^ {2} \left[
\left({\frac {\partial f}{\partial z}}\right )
\left(\frac {\partial f}{\partial y}\right)
-
\frac{1}{2}\,(1-f){\frac {\partial ^{2}f}{\partial
y\partial z}}
\right]
 \,,
\\
G_{\hat{y}\hat{y}}
&=&
v^2 \left[ \frac{1}{4}\, \left({\frac
{\partial f}{\partial y}} \right )^{2}
-
\frac{1}{4}\,\left({\frac
{\partial f}{\partial x}} \right )^{2}
-
\left ({\frac {\partial
f}{\partial z}}\right )^{2}
+
(1-f){\frac {\partial ^{2}f}{\partial {z}^{2}}} \right]  \,,
\\
G_{\hat{z}\hat{z}}
&=&
-\frac{3}{4}\,{v}^{2}\left[
\left (\frac{\partial f}{\partial x} \right )^2
+
\left (\frac {\partial f}{\partial y} \right )^2
\right]   \,.
\end{eqnarray}

\subsection{Alcubierre form function}

Taking into account Alcubierre's choice of the form function,
\emph{i.e.}, $f(r)$ with $r=\sqrt{x^2+y^2+[z-z_0(t)]^2}$, the Einstein
tensor components of the warp drive spacetime, in cartesian
coordinates, in an orthonormal basis, $G_{\hat{\mu}\hat{\nu}}$, take
the following form
\begin{eqnarray}
G_{\hat{t}\hat{t}} &=& - \frac{v^2}{4}\, \frac{(x^2+y^2)}{r^2}
\left(\frac{\d f}{\d r} \right)^2 \,,
\\
G_{\hat{t}\hat{x}} &=& - \frac{v}{2}\, \frac{x(z-z_0(t))}{r^2}
\left(\frac{\d^2f}{\d r^2}-\frac{1}{r}\frac{\d f}{\d r} \right)
\,,
\\
G_{\hat{t}\hat{y}} &= &- \frac{v}{2}\, \frac{y(z-z_0(t))}{r^2}
\left(\frac{\d^2f}{\d r^2}-\frac{1}{r}\frac{\d f}{\d r} \right)
\,,
\\
G_{\hat{t}\hat{z}} &=& \frac{v}{2}\, \left[\frac{x^2+y^2}{r^2}\,
\frac{\d^2f}{\d r^2}+ \frac{x^2+y^2+2(z-z_0(t))^2}{r^3}\, \frac{\d
f}{\d r} \right] \,,
\\
G_{\hat{x}\hat{x}} &=&
\frac{v^2}{4}\,\frac{x^2-y^2-4(z-z_0(t))^2}{r^2}\, \left(\frac{\d
f}{\d r}\right)^2 +v^2(1-f)\, \left[\frac{(z-z_0(t))^2}{r^2}\,
\frac{\d^2f}{\d r^2}+ \frac{x^2+y^2}{r^3}\, \frac{\d f}{\d r}
\right] \,,
\\
G_{\hat{x}\hat{y}} &=& \frac{v^2}{2}\,\frac{xy}{r^2}\,
\left(\frac{\d f}{\d r}\right)^2 \,,
\\
G_{\hat{x}\hat{z}} &=& \frac{v^2}{2}\,\frac{x(z-z_0(t))}{r^2}\,
\left[2\left(\frac{\d f}{\d r}\right)^2-(1-f)\left(\frac{\d^2f}{\d
r^2}- \frac{1}{r}\, \frac{\d f}{\d r} \right)\right] \,,
\\
G_{\hat{y}\hat{z}} &=& \frac{v^2}{2}\,\frac{y(z-z_0(t))}{r^2}\,
\left[2\left(\frac{\d f}{\d r}\right)^2-(1-f)\left(\frac{\d^2f}{\d
r^2}- \frac{1}{r}\, \frac{\d f}{\d r} \right)\right] \,,
\\
G_{\hat{y}\hat{y}} &=&
\frac{v^2}{4}\,\frac{y^2-x^2-4(z-z_0(t))^2}{r^2}\, \left(\frac{\d
f}{\d r}\right)^2 +v^2(1-f)\, \left[\frac{(z-z_0(t))^2}{r^2}\,
\frac{\d^2f}{\d r^2}+ \frac{x^2+y^2}{r^3}\, \frac{\d f}{\d r}
\right] \,,
\\
G_{\hat{z}\hat{z}} &=& -\frac{3}{4}\,{v}^{2} \,
\frac{(x^2+y^2)}{r^2} \left(\frac{\d f}{\d r} \right)^2 \,.
\end{eqnarray}

\subsection{Linearized Einstein tensor for the Alcubierre warp drive}

Linearizing in the warp velocity is very easy. The only nonzero
components are
\begin{eqnarray}
G_{\hat{t}\hat{x}}
&=&
-\frac{1}{2}\,v{\frac {\partial^{2}\,f}{\partial x \partial z}}  \,,
\\
G_{\hat{t}\hat{y}}
&=
&-\frac{1}{2}\,v{\frac {\partial^{2}\,f}{\partial y \partial z}}  \,,
\\
G_{\hat{t}\hat{z}} &=& \frac{1}{2}\,v\left( {\frac {\partial
^{2}f}{\partial {x}^{2}}} +\frac {\partial ^{2}f}{\partial
{y}^{2}} \right)  \,.
\end{eqnarray}
All other components are $O(v^2)$.

For the Alcubierre form function $f(r)$ we find that at linear order
in warp bubble velocity
\begin{eqnarray}
G_{\hat{t}\hat{x}} &=& - \frac{v}{2}\, \frac{x(z-z_0(t))}{r^2}
\left(\frac{\d^2f}{\d r^2}-\frac{1}{r}\frac{\d f}{\d r} \right)
\,,
\\
G_{\hat{t}\hat{y}} &= &- \frac{v}{2}\, \frac{y(z-z_0(t))}{r^2}
\left(\frac{\d^2f}{\d r^2}-\frac{1}{r}\frac{\d f}{\d r} \right)
\,,
\\
G_{\hat{t}\hat{z}} &=& \frac{v}{2}\, \left[\frac{x^2+y^2}{r^2}\,
\frac{\d^2f}{\d r^2}+ \frac{x^2+y^2+2(z-z_0(t))^2}{r^3}\, \frac{\d
f}{\d r} \right] \,.
\end{eqnarray}
All other components are $O(v^2)$.

\section{The Einstein tensor for the Nat\'{a}rio warp drive}
\label{A:natario}

The Einstein tensor for the Nat\'{a}rio warp drive is relatively
easy to calculate once one realizes that it can be dealt with
within the framework of ``shift-only'' spacetimes.  The extrinsic
curvature of the constant-$t$ time slices is
\begin{equation}
K_{ij} = \beta_{(i;j)} =
{1\over2} \left[\partial_i \beta_j + \partial_j \beta_i \right]
\end{equation}
and the no expansion condition implies
\begin{equation}
K = \hbox{Tr}({\bf K}) = 0
\end{equation}
so the trace of extrinsic curvature is zero. The intrinsic curvature
of the constant-$t$ time slices is by construction zero (since they are
flat). It is useful to introduce a vorticity tensor
\begin{equation}
\Omega_{ij} = - \beta_{[i,j]} =
{1\over2} \left[\partial_i \beta_j - \partial_j \beta_i \right]
\end{equation}
and then a tedious but straightforward computation (which follows a
variant of the Gauss--Codazzi decomposition) yields
\begin{eqnarray}
\label{E:Riemann1}
R_{ijkl} &=& K_{ik} K_{jl}-K_{il} K_{jk} \,,
\\
R_{tijk} &=& -\partial_i \Omega_{jk}
+\beta_l \left( K_{kl} K_{ij} - K_{jl}K_{ik}\right)\,,
\\
R_{titj} &=& -\partial_t K_{ij} + \left({\bm K}{\bm \Omega}
+ {\bm \Omega}{\bm K}\right)_{ij}
-\left({\bm K}^2\right)_{ij}
-\beta_k \beta_{k,ij}
+\beta_k \beta_l \left( K_{kl} K_{ij} - K_{jk}K_{il}\right)\,.
\end{eqnarray}
Here we have defined
\begin{equation}
\left({\bm K}{\bm \Omega}+ {\bm \Omega}{\bm K}\right)_{ij}
\equiv K_{ik}\Omega_{kj}+ \Omega_{ik} K_{kj},
\end{equation}
and similarly
\begin{equation}
\left({\bm K}^2\right)_{ij} \equiv K_{ik} K_{kj}.
\end{equation}
The computation reported here for the Nat\'{a}rio warp drive is
identical in spirit (and differs in only notation and minor
technical details) to the computation of the ``acoustic'' Riemann
tensor for the ``acoustic geometry'' reported in
\cite{Fischer1,Fischer2}. Specifically, when sound waves propagate
in a constant-density constant speed-of-sound background fluid
flow the acoustic metric is identical in form to Nat\'{a}rio's
metric.

\def\ii{{\hat\imath}}
\def\jj{{\hat\jmath}}
\def\kk{{\hat k}}
\def\lll{{\hat l}}
\def\tt{{\hat t}}
\def\xx{{\hat x}}
\def\yy{{\hat y}}
\def\zz{{\hat z}}

The Einstein tensor is now (in an orthonormal basis)
\begin{eqnarray}
G_{\tt\tt} &=&
-\frac12 {\rm Tr}({\bm K}^2)\,,
\\
G_{\tt\ii} &=&
\frac12 \Delta \beta_i\,,
\\
G_{\ii\jj} &=&
\frac{\d}{\d t} K_{ij}
-\frac12 \delta_{ij} {\rm Tr}({\bm K}^2)
-\left({\bm K}{\bm \Omega}+ {\bm \Omega}{\bm K}\right)_{ij},
\end{eqnarray}
where $\d/\d t$ denotes the usual advective derivative
\begin{equation}
{\d\over\d t} = {\partial\over\partial t} +
\beta^i\; {\partial\over \partial x^i}
\end{equation}
In view of the fact that for the Nat\'{a}rio spacetime all the
time dependence comes from motion of the warp bubble, so that all
fields have a space-time dependence of the form
$\beta(x,y,z-z_0(t))$, we can deduce
\begin{equation}
{\d\over\d t} \to - v(t) {\partial\over \partial z} +
\beta^i\; {\partial\over \partial x^i}
= - \left[  v(t) \hat z - \vec \beta \right] \cdot \nabla
\end{equation}
so that
\begin{equation}
G_{\ii\jj} \to
 - \left[  v(t) \; \hat z - \vec \beta \right] \cdot \nabla K_{ij}
-\frac12 \delta_{ij} {\rm Tr}({\bm K}^2) -\left({\bm K}{\bm
\Omega}+ {\bm \Omega}{\bm K}\right)_{ij} \,.
\end{equation}

If we now linearize, then the only surviving term in the Einstein
tensor is $G_{\tt\ii} = \frac12 \Delta \beta_i$ as all other terms are
$O(v^2)$.

A particularly simple special case of the Nat\'{a}rio spacetime is
obtained if the shift vector $\beta$ is taken to be irrotational
as well as being divergenceless~\cite{Fischer2} In this special
case $\Omega\to0$ and also $\Delta \beta_i \to 0$, so that (in an
orthonormal basis)
\begin{eqnarray}
G_{\tt\tt} &=&
-\frac12 {\rm Tr}({\bm K}^2)\,,
\\
G_{\tt\ii} &=& 0\,,
\\
G_{\ii\jj} &=&
 - \left[  v(t) \; \hat z - \vec \beta \right] \cdot\nabla K_{ij}
-\frac12 \delta_{ij} {\rm Tr}({\bm K}^2)  \,.
\end{eqnarray}
When linearizing this special case we find that \emph{all} components
of the Einstein tensor are $O(v^2)$.

\section{Linearized Gravity}
\label{A:linear}

For a weak gravitational field, linearized around flat Minkowski
spacetime, we can write the spacetime metric
as~\cite{Misner,Wald,Schutz}
\begin{equation}
g_{\mu\nu}=\eta_{\mu\nu}+h_{\mu\nu}  \label{linearmetric}  \,,
\end{equation}
with $h_{\mu\nu}\ll 1$. It is convenient to raise and lower tensor
indices with $\eta^{\mu\nu}$ and $\eta_{\mu\nu}$, respectively, rather
than with $g^{\mu\nu}$ and $g_{\mu\nu}$. We shall adopt this
notational convention, keeping in mind that the tensor $g^{\mu\nu}$
still denotes the inverse metric, and not $\eta^{\mu\alpha}\,
\eta^{\nu\beta}\, g_{\alpha\beta}$. It should also be noted that in
linear approximation we have $g^{\mu\nu} = \eta^{\mu\nu} -
h^{\mu\nu}$, since the composition of $g_{\mu\nu}$ and $g^{\mu\nu}$
differ from the identity operator only by terms quadratic in
$h_{\mu\nu}$.

The linearized Einstein equation can be obtained in the
straightforward manner as follows. Adopting the form of equation
(\ref{linearmetric}) for the metric components, the resulting
connection coefficients, when linearized in the metric perturbation
$h_{\mu\nu}$, yield
\begin{eqnarray}
\Gamma^{\mu}{}_{\alpha\beta}
&=&
\frac{1}{2}\eta^{\mu\nu}\left(h_{\alpha\nu,\beta}
+h_{\beta\nu,\alpha}-h_{\alpha\beta,\nu} \right)
\\
\nonumber
&=&
\frac{1}{2}\left(h^{\mu}{}_{\alpha,\beta}
+h^{\mu}{}_{\beta,\alpha}-h_{\alpha\beta,}{}^{\mu}\right)
\label{linearChristoffel}  \,.
\end{eqnarray}
A similar linearization of the Ricci tensor, $R_{\mu\nu} =
R^{\alpha}{}_{\mu\alpha\nu}$, given as a function of the Christoffel
symbols, i.e.,
\begin{equation}
R_{\mu\nu}
=
\Gamma^{\alpha}{}_{\mu\nu,\alpha}
-
\Gamma^{\alpha}{}_{\mu\alpha,\nu}
+
\Gamma^{\alpha}{}_{\beta\alpha} \Gamma^{\beta}{}_{\mu\nu}
-
\Gamma^{\alpha}{}_{\beta\nu} \Gamma^{\beta}{}_{\mu\alpha} \,,
\end{equation}
results in the following approximation
\begin{equation}
R_{\mu\nu}
=
\Gamma^{\alpha}{}_{\mu\nu,\alpha}
-
\Gamma^{\alpha}{}_{\mu\alpha,\nu}
=
\frac{1}{2}\left(h^{\alpha}{}_{\mu\nu,\alpha}
+
h^{\alpha}{}_{\nu,\mu\alpha}
-
h_{\mu\nu,\alpha}{}^{\alpha}
\right) \label{linearRicci}  \,,
\end{equation}
where $h = h^{\alpha}{}_{\alpha} =\eta^{\alpha\beta}\,
h_{\beta\alpha}$ is the trace of $h_{\alpha\beta}$. Applying a further
contraction to discover the Ricci scalar, $R = g^{\mu\nu}\;
R_{\mu\nu}\simeq \eta^{\mu\nu} R_{\mu\nu}$, the Einstein tensor to
linear order is given by
\begin{equation}
G_{\mu\nu}
=
\frac{1}{2}\left[h_{\mu\alpha,\nu}{}^{\alpha}
+
h_{\nu\alpha,\mu}{}^{\alpha}
-
h_{\mu\nu,\alpha}{}^{\alpha}
-
h_{,\mu\nu}-\eta_{\mu\nu}\left(h_{\alpha\beta,}{}^{\alpha\beta}
-
h_{,\beta}{}^{\beta}\right)\right]
\label{firstlinearEinstein}\,,
\end{equation}
Equation (\ref{firstlinearEinstein}) can be simplified by defining the
{\it trace reverse} of $h_{\alpha\beta}$, given by
\begin{equation}
\overline{h}_{\alpha\beta}
=
h_{\alpha\beta}-\frac{1}{2}\eta_{\alpha\beta}\,h
\label{tracereverse} \,,
\end{equation}
with $\overline{h}=\overline{h}^{\alpha}{}_{\alpha}=-h$.  Therefore,
in terms of $\overline{h}_{\alpha\beta}$, the linearized Einstein
tensor reads
\begin{equation}
G_{\alpha\beta}
=
-
\frac{1}{2}\left[\overline{h}_{\alpha\beta,\mu}{}^{\mu}
+
\eta_{\alpha\beta} \overline{h}_{\mu\nu,}{}^{\mu\nu}
-
\overline{h}_{\alpha\mu,\beta}{}^{\mu}
-
\overline{h}_{\beta\mu,\alpha}{}^{\mu}
+
O\left(\overline{h}_{\alpha\beta}^2\right)\right]
\label{linearEinstein} \,,
\end{equation}
and the linearized Einstein equation is found to be
\begin{equation}
-
\overline{h}_{\alpha\beta,\mu}{}^{\mu}
-
\eta_{\alpha\beta}
\overline{h}_{\mu\nu,}{}^{\mu\nu}
+
\overline{h}_{\alpha\mu,\beta}{}^{\mu}
+
\overline{h}_{\beta\mu,\alpha}{}^{\mu}
=16\pi\, T_{\alpha\beta}  \,.
\end{equation}

Without loss of generality, one can impose the gauge condition,
i.e.,
\begin{equation}
\overline{h}^{\mu\nu}{}_{,\nu}=0 \label{gaugecondition} \,,
\end{equation}
which is similar to the tensor analogue of the Lorenz gauge
$A^{\mu}{}_{,\mu}=0$ of electromagnetism \cite{Lorenz}. The Einstein
equation then simplifies to:
\begin{equation}
-\overline{h}_{\alpha\beta,\mu}{}^{\mu}=16\pi\,
T_{\alpha\beta}
\label{linearizedfieldequation}   \,.
\end{equation}
The gauge condition, equation (\ref{gaugecondition}), the field
equation, equation (\ref{linearizedfieldequation}), and the
definition of the metric
\begin{equation}
g_{\mu\nu}
=
\eta_{\mu\nu}+h_{\mu\nu}
=
\eta_{\mu\nu}+\overline{h}_{\mu\nu}
-\frac{1}{2} \,\eta_{\mu\nu}\,\overline{h}
\end{equation}
are the fundamental equations of the linearized theory of gravity
written in the Lorenz gauge \cite{Lorenz}.



\end{document}